# Measures of 62 Southern Pairs.

Matthew James[1], Rod letchford[2], Graeme L. White[2], Meg Emery[3] and Stephen Bosi[1].

1. University of New England, Armidale, NSW, Australia; m.b.james27@gmail.com; sbosi@une.edu.au
2. Centre for Astronomy, University of Southern Queensland, Toowoomba, QLD, Australia; Rod.Letchford@usq.edu.au; graemewhiteau@gmail.com
3. Kildare Catholic College, Wagga Wagga, NSW, Australia; megcolemery@gmail.com

**Abstract:** We report lucky imaging observations of 62 pairs at mid-southern declinations sourced from the WDS with separations larger than 4 arc seconds and magnitude less than 10. The measures comprise separations and PA calibrated against Alpha Centauri AB and drift scans, presented as weighted means of these two calibration methods, with formal internal uncertainties $\delta\rho = 80$ mas and $\delta PA = 0.056°$.

We also compare our measures against 1) extrapolated historic measures, 2) GAIA DR2 data and 3) measures determined from HIPPARCOS and GAIA observations. Our best estimate of our bias against these 3 databases are $\rho \approx 10 \pm 30$ mas and $PA \approx 0.04 \pm 0.08°$. These formal uncertainties are consistent with the internal uncertainties of $\delta\rho = 80$ mas and $\delta PA = 0.056°$.

We also report Rectilinear Elements for 61 pairs, Grade 5 Orbital Elements for 5 pairs and suggest 5 pairs as optical doubles (4 of which are new).

## 1. Introduction.

This companion paper to (James, *et al*. 2019), reports lucky imaging (Fried, 1977) observations of 62 southern pairs. These data comprise separations, $\rho$, and position angles, PA, determined using two different methods of calibration; the video-drift method and calibration against published orbital parameters of Alpha Centauri AB (α Cen AB). Results are compared with extrapolations of historic observations and micro-arcsecond precision positions from the GAIA DR2 (Brown, *et al*. 2018, Prusti, *et al*. 2016) and HIPPARCOS databases. A rectilinear analysis is presented and, where possible, orbital parameters estimated.

## 2. Selection of Pairs.

The 62 pairs in this paper were sourced from the Washington Double Star Catolog (WDS, Mason, *et al*. 2001) with separations limited to those larger than 4 arc seconds and secondary magnitude brighter than 10. These limits were imposed by the instrument's capabilities. All are at mid-southern declinations such that they would be close to the zenith during observation. The location of the pairs on the celestial sphere is shown in Figure 1.

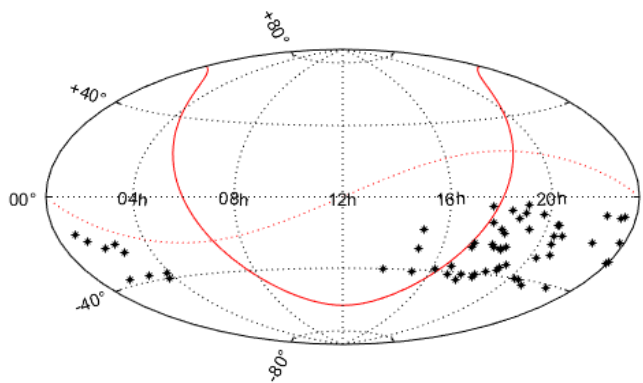

*Figure 1*: Location of the 62 pairs

### 3. Observations.

These observations of the 62 pairs follow Paper 1 (James, *et al*. 2019). In brief, observations were made with the Webster Celestron 14-inch telescope of The University of New England (UNE) Kirby observatory using an ASI120MM-S camera fitted with a Wratten #25 (red) filter. The capture and reduction software used was *SharpCaps* version 3.1 (see Paper 1) and *Reduc* version 5.36 (provided by Florent Losse).

Two observing techniques were used. The first was traditional lucky imaging calibrated against $\rho$ and PA of α Cen AB where precise values of $\rho$ and PA were determined from orbital parameters presented by Pourbaix and Boffin (2016).

The second method was the video-drift method of Nugent & Iverson (2011) where the telescope is stationary and where $\rho$ (which is derived from the image scale of the detector) is determined from the 'angular velocity' of the image movement across the camera detector. The PA is determined by the 'direction' of the movement of the image over the chip.

These two observational techniques were adopted with the intent of finding the best method of calibrating instruments similar in capability to the ones used in this investigation. This is discussed in our follow-up paper (James, *et al*. 2020b).

### 4. Astrometric Measures.

The measured $\rho$ and PA for the 62 pairs are given in Table 1. Columns 1 and 2 give the WDS designation and discovery code respectively. Column 3 gives the epoch of the new observations reported here.

Columns 3 and 4 give the weighted separation of the two observational techniques in arcsec and the formal uncertainty in $\rho$, and columns 5 and 6 give the weighted PA in degrees and the formal uncertainty in PA. The average uncertainty of the weighted mean separation is 0.080 arcsec and the average weighted mean uncertainty in PA is 0.056 deg.

*Table 1*: Weighted Average Measures of Drift and Lucky Imaging (Tracking) Observations for the 62 Pairs.

| WDS | DISC | Epoch | Sep (arcsec) | SEM Sep (arcsec) | PA (deg) | SEM PA (deg) |
|---|---|---|---|---|---|---|
| 00582-1541 | S 390 | 2018.696 | 6.452 | 0.015 | 216.290 | 0.059 |
| 01225-1905 | HJ 2043 | 2018.712 | 5.115 | 0.020 | 72.134 | 0.309 |
| 01397-3728 | HJ 3452 | 2018.712 | 19.610 | 0.082 | 277.769 | 0.024 |
| 01590-2255 | H 2 58AB | 2018.696 | 8.711 | 0.022 | 302.246 | 0.074 |
| 02360-2124 | HJ 3511 | 2018.712 | 14.702 | 0.042 | 98.232 | 0.058 |
| 02442-2530 | BSO 1AB | 2018.712 | 12.529 | 0.042 | 192.518 | 0.062 |
| 02486-3724 | HJ 3532 | 2018.712 | 5.320 | 0.019 | 144.265 | 0.054 |
| 03398-4022 | DUN 15 | 2018.712 | 7.681 | 0.026 | 328.043 | 0.048 |
| 03486-3737 | DUN 16 | 2018.712 | 8.409 | 0.042 | 217.178 | 0.044 |
| 13493-4031 | DUN 146 | 2018.562 | 68.511 | 0.306 | 86.642 | 0.019 |
| 15036-2751 | HJ 4727AB | 2018.658 | 7.346 | 0.020 | 221.429 | 0.107 |
| 15045-1754 | S 665 | 2018.658 | 25.073 | 0.078 | 90.534 | 0.019 |
| 15103-4100 | COO 178 | 2018.658 | 4.872 | 0.030 | 74.720 | 0.216 |
| 16086-3906 | DUN 199AC | 2018.658 | 44.441 | 0.107 | 183.572 | 0.019 |
| 16482-3653 | DUN 209AB | 2018.562 | 24.011 | 0.061 | 138.274 | 0.039 |
| 16540-4148 | JC 23AF | 2018.562 | 56.702 | 0.496 | 19.114 | 0.017 |
| 17153-2636 | SHJ 243AB | 2018.696 | 5.140 | 0.011 | 139.764 | 0.087 |
| 17180-2417 | H 3 25 | 2018.696 | 9.988 | 0.027 | 353.349 | 0.095 |
| 17290-4358 | DUN 217 | 2018.696 | 13.427 | 0.030 | 168.274 | 0.048 |
| 17317-4102 | ARY 114AC | 2018.562 | 64.868 | 0.169 | 250.599 | 0.009 |
| 17398-0458 | STF2191AB | 2018.696 | 26.350 | 0.063 | 266.705 | 0.030 |
| 17464-1318 | STF2204 | 2018.696 | 14.509 | 0.051 | 24.027 | 0.053 |
| 18026-2415 | ARG 31AC | 2018.614 | 35.600 | 0.089 | 26.296 | 0.013 |
| 18064-4145 | HJ 5011 | 2018.614 | 28.143 | 0.075 | 344.074 | 0.030 |
| 18089-2528 | WNO 21 | 2018.614 | 13.484 | 0.048 | 64.327 | 0.069 |
| 18106-1645 | S 700AB | 2018.696 | 18.812 | 0.101 | 291.235 | 0.151 |
| 18106-1645 | S 700AC | 2018.696 | 28.618 | 0.088 | 352.874 | 0.032 |
| 18108-4026 | HJ 5023 | 2018.696 | 8.654 | 0.029 | 275.919 | 0.050 |
| 18187-1837 | SHJ 264AB,C | 2018.696 | 17.230 | 0.050 | 51.048 | 0.039 |
| 18247-0636 | STF2313 | 2018.614 | 5.762 | 0.019 | 195.693 | 0.072 |
| 18281-2645 | ARY 126AC | 2018.600 | 54.054 | 0.125 | 135.204 | 0.051 |
| 18290-2635 | WNO 6 | 2018.600 | 41.804 | 0.099 | 182.311 | 0.049 |
| 18334-3844 | DUN 222 | 2018.562 | 20.470 | 0.053 | 358.654 | 0.023 |
| 18399-2531 | ARG 32AB | 2018.616 | 7.032 | 0.019 | 219.503 | 0.064 |
| 18399-2531 | ARG 32AC | 2018.616 | 53.217 | 0.126 | 284.941 | 0.009 |
| 18459-1030 | STF2373 | 2018.616 | 4.117 | 0.012 | 337.020 | 0.034 |
| 19006-0807 | STF2425 | 2018.616 | 29.534 | 0.086 | 177.100 | 0.027 |
| 19011-3704 | BSO 14AB | 2018.559 | 12.825 | 0.037 | 280.033 | 0.051 |
| 19027-3606 | HJ 5080 | 2018.616 | 5.484 | 0.011 | 245.883 | 0.066 |
| 19050-0402 | SHJ 286 | 2018.616 | 39.705 | 0.157 | 209.937 | 0.028 |
| 19127-3351 | HJ 5094AB | 2018.600 | 31.571 | 0.104 | 181.613 | 0.048 |
| 19177-1558 | S 715 | 2018.600 | 8.380 | 0.022 | 16.967 | 0.054 |
| 19181-1557 | S 716 | 2018.600 | 4.969 | 0.012 | 196.095 | 0.078 |
| 19431-0818 | STF45AB | 2018.636 | 96.268 | 0.460 | 145.488 | 0.015 |
| 20178-4011 | DUN 230 | 2018.562 | 9.675 | 0.036 | 117.438 | 0.065 |
| 20205-2912 | HJ 5188AC | 2018.712 | 27.266 | 0.084 | 320.898 | 0.025 |
| 20284-1309 | STF2683 | 2018.712 | 22.795 | 0.100 | 66.820 | 0.067 |
| 20299-1835 | SHJ 324 | 2018.636 | 21.947 | 0.083 | 238.443 | 0.052 |
| 20322-2209 | HJ 2973AB | 2018.712 | 39.386 | 0.119 | 128.953 | 0.018 |
| 20338-4033 | JC 18AB | 2018.562 | 4.413 | 0.014 | 223.593 | 0.090 |
| 20484-1812 | S 763AB | 2018.636 | 15.605 | 0.051 | 293.786 | 0.024 |
| 20501-2722 | HJ 5226 | 2018.712 | 18.658 | 0.065 | 67.357 | 0.038 |
| 21022-4300 | DUN 236 | 2018.562 | 57.422 | 0.265 | 73.193 | 0.018 |
| 22246-4127 | JC 19AB | 2018.562 | 14.790 | 0.045 | 58.811 | 0.038 |
| 22258-2014 | S 808AB | 2018.658 | 6.869 | 0.020 | 152.910 | 0.091 |
| 22305-0807 | STF2913 | 2018.658 | 8.067 | 0.025 | 327.936 | 0.054 |
| 23141-0855 | STF2993AB | 2018.658 | 25.068 | 0.061 | 175.134 | 0.040 |
| 23141-0855 | S 826AC | 2018.658 | 79.127 | 0.336 | 131.875 | 0.020 |
| 23238-0828 | STF3008 | 2018.636 | 7.053 | 0.028 | 146.790 | 0.079 |
| 23460-1841 | H 2 | 2018.614 | 6.950 | 0.021 | 135.798 | 0.090 |
| 23544-2703 | LAL 192 | 2018.614 | 6.455 | 0.022 | 272.184 | 0.066 |
| 23595-2631 | LAL 193 | 2018.614 | 10.373 | 0.031 | 169.436 | 0.039 |

## 5. Historic Data.

The historical data for the 62 pairs was requested from Brian Mason of the United States Naval Observatory (UNSO).

It is worth reminding the reader that, if not diligent, false conclusions can be drawn from otherwise seemingly clear trends in data. Figure 2 shows an instance of what at first might seem to be evidence for orbital motion. However, such a conclusion would be spurious. For the pair WDS 23238-0828, the left plots are of $\rho$ and PA with Epoch. Both plots show strong curvature which might be taken (uncritically) as evidence that the secondary is in orbit around the primary. The true sky motion of the pair is revealed in the right-side rectilinear plot which shows the true straight-line motion of the secondary with respect to the primary.

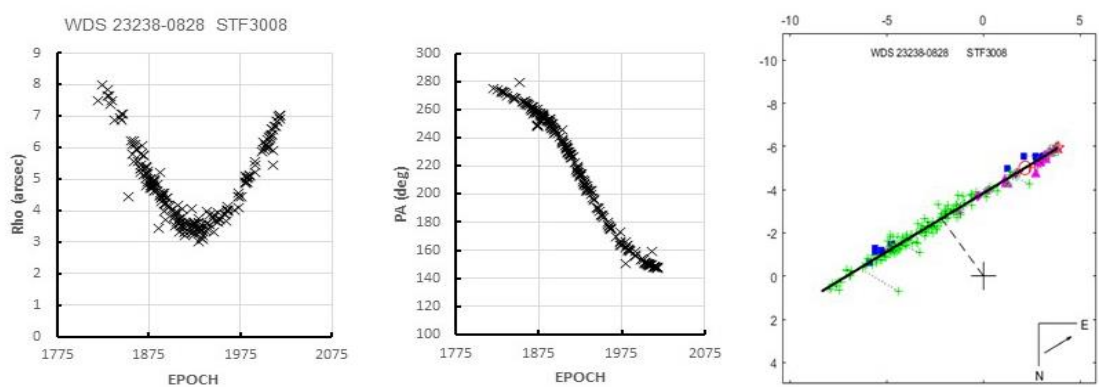

*Figure 2*: (Left to right) $\rho$ with Epoch, PA with Epoch and Rectilinear Plot for pair WDS 23238-0828.

## 6. Rectilinear Motion for 62 Pairs.

The rectilinear motion of the 62 pairs is presented in Table A5 and Figure A5 (in the Appendix). The Rectilinear Elements x0, xa, y0, ya, t0, θ0 and $\rho$0 are defined in Letchford, White and Ernest, 2018a. In this paper the value of t0 differs in that it is set equal to the mean of observational epochs (2018.6 ± 0.15 - over a period of 56 days). The broad straight line is the proper motion trajectory of the secondary, relative to the primary, based on the HIPPARCOS and GAIA positions only.

Additionally, no available HIPPARCOS or ASCC data were available on component B of the pair WDS 15103-4100 (COO 178) exists; thus, it is not included in the rectilinear plots (Figure 5) or rectilinear element table (Table 5).

Quick scrutiny of the rectilinear plots of Figure 5 shows:

- The obvious erroneous observation among the historic data set. These have been rejected from the computation of the extrapolated historic data.

- For some observations where the two stars are of very similar magnitude, an obvious misidentification of the primary has occurred, and the recorded PA is in error by 180 degrees. They are (magnitudes from SIMBAD):
  i) WDS 00582-1514 (HD 5659) where the V magnitudes of the components are 7.712 and 7.804.
  ii) WDS 17153-2636 (36 Oph) where V = 5.03 and 5.08.
  iii) WDS 19177-1558 (HD 180562) where V = 7.038 and 7.87.
  iv) WDS 19181-1557 (HD 180695) where V = 7.712 and 7.804.

Irrespective of these data misadventures, the rectilinear plots are of three types. The first type is a curved path of a companion in a short period orbit. In Figure 5 (Appendix) WDS 17153-2636 (36 Oph) shows such a curved path and an orbit of period 507 ± 18 years is determined by us in Table 6 & Figure 6 (Appendix).

### 6.1. Five Optical Doubles.

A clear straight-line trajectory of the secondary relative to a stationary primary is expected when a pair is an optical double and large differences in parallax are an indication of non-gravitational systems. Examples of linear trajectory rectilinear plots in Figure 5 are (parallaxes are from GAIA DR2, units are milli-arcseconds (mas));

- WDS 19127-3351, Parallax of the primary is 1.97 ± 0.11 and the parallax of the secondary is 13.88 ± 0.13.
- WDS 22246-4127, Parallax of the primary is 15.53 ± 0.05, parallax of the secondary is 7.29 ± 0.07.
- WDS 23141-0855, Parallax of the primary is 24.55 ± 0.05, parallax of the secondary is 11.38 ± 0.05.
- WDS 23460-1841, Parallax of the primary is 16.2 ± 0.2, the parallax of the secondary is 20.4 ± 0.4.
- WDS 23238-0828, Parallax of the primary is 5.45 ± 0.04 and the parallax of the secondary is 9.43 ± 0.04. This is a confirmed optical pair (Peirce, 1882; MacEvoy, 2010).

On the basis of the Rectilinear plots of Figure 5, and the very different parallaxes, we confidently categorise 4 of these pairs as optical doubles and confirm the fifth (WDS 23238-0828) as an optical double.

### 7. Orbits for 5 Pairs.

Following the technique presented in Letchford, White and Ernest, 2018b, table A6 presents Grade 5 Orbital Elements for five pairs. All pairs reported here display very short arcs. Column headings in Table A6 are described in Letchford, White and Ernest, 2018b. For this analysis, all historic data is considered and unweighted. Figure 6 (Appendix) shows these best fit and iterative orbits. WDS 17153-2636 (36 Oph) has an orbit of period 507 ± 18 years. The WDS Sixth Orbit Catalog reports a 470.9-year period from Irwin, Yang and Walker (1996), which is based on the period of 548.7 years derived by Brosche (1960). Our value of 507 ± 18 years in consistent with these prior computations.

A literature search for the binary status of 5 pairs found 3 as 'known binaries' (referenced in ADS and SIMBAD).

- WDS 01590-2255 is a known binary
- WDS 02360-2124 is a known binary, however, has only been suggested to be a bound system recently by Andrews, Chaname & Agueros, 2017.
- WDS 17153-2636 is a known binary
- WDS 18247-0636 is a known binary
- WDS 23595-2636 is an unknown binary

Regarding the unknown pair WDS 23590-2636, a grade 5 orbital is presented in Table 6 and Figure 6. Within the confidence of our results we nominate WDS 23590-2636 as a binary system.

## 8. Precision, Bias and Comparison with Other Data Sets.
### 8.1 Comparison with History Data.

Following Paper 1, linear extrapolations of the historic data were derived and the parameters A, B, C, and D were determined. These define the best linear fit to the trends in the data.

The parameters A-D were used to determine the predicted $\rho$ and PA at the epoch of our observation and these values are given in Table 3 (Appendix).

There is good agreement between the weighted measures reported here and the measures obtained from extrapolation. The mean bias in $\rho$ is -0.009 ± 0.027 arcseconds, and the mean bias in PA is -0.098 ± 0.078 degree; the bias is defined as the difference, this work (TW) minus historic data.

Figure 3 shows the difference in $\rho$ and PA for 57 of the 62 pairs measures of this work with those determined from linear fits of the historic data. Pairs with a poor historic linear fit were not appropriate for comparison and were treated as outliers, the criterion being that $\rho$ or PA differs by 2 standard deviations or more, from the mean of the differences. The following pairs 18334-3844 (DUN 222), 19127-3351 (HJ 5094AB), 22246-4127(JC 19AB), 23141-0855 (S 826AC), 23238-0828 (STF3008) were treated as outliers here.

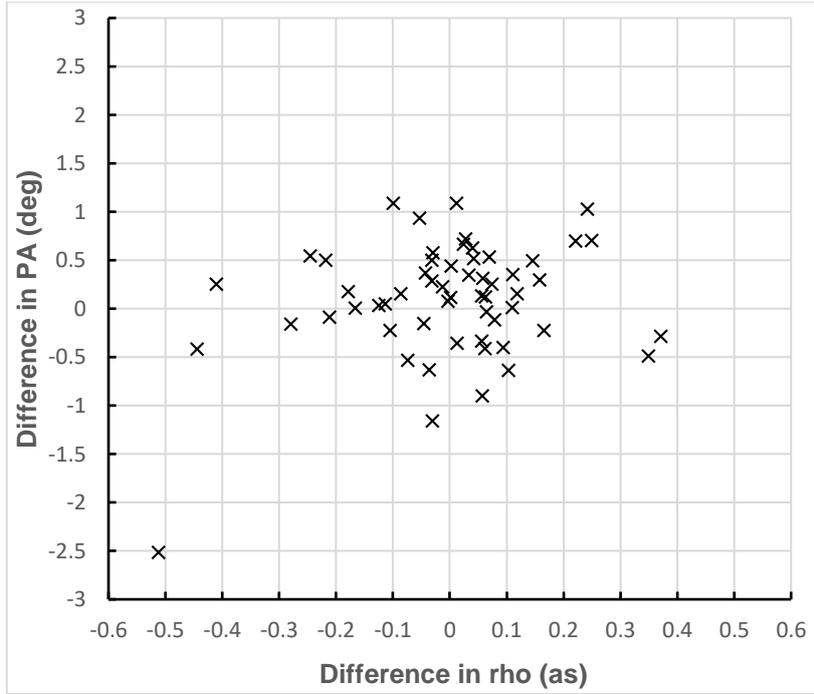

*Figure 3*: *The comparison of measures (extrapolation of historic data – This paper (TW)). The mean bias in ρ is -0.009 ± 0.027 arcseconds, and the mean bias in PA is 0.098 ± 0.078 degree.*

## 8.2 Comparison with GAIA DR2 Data.

Once more following Paper 1, Table A3 (Appendix) also gives the comparison of the measures in Table 1 (this work) with precision values from GAIA DR2. All measures are compared at the epoch of observation (~2018.6).

Again, there is good agreement between the weighted measures reported here and the measures obtained from GAIA DR2. The mean bias in $\rho$ is 0.005 ± 0.021 arcseconds, and the mean bias in PA is 0.129 ± 0.094 degree. Again, here bias is defined as the difference, TW minus GAIA DR2 data.

Figure 4 shows the difference (for 58 of the 62 pairs measures) between this work and those determined from GAIA DR2. Pairs outside 2 standard deviations of the mean difference were considered outliers and omitted from the plot. We suspect that our work may have isolated discrepancies in the DR2. The following pairs 18334-3844 (DUN 222), 22246-4127 (JC 19AB), 23141-0855 (S 826AC), 23238-0828 (STF3008) were treated as outliers.

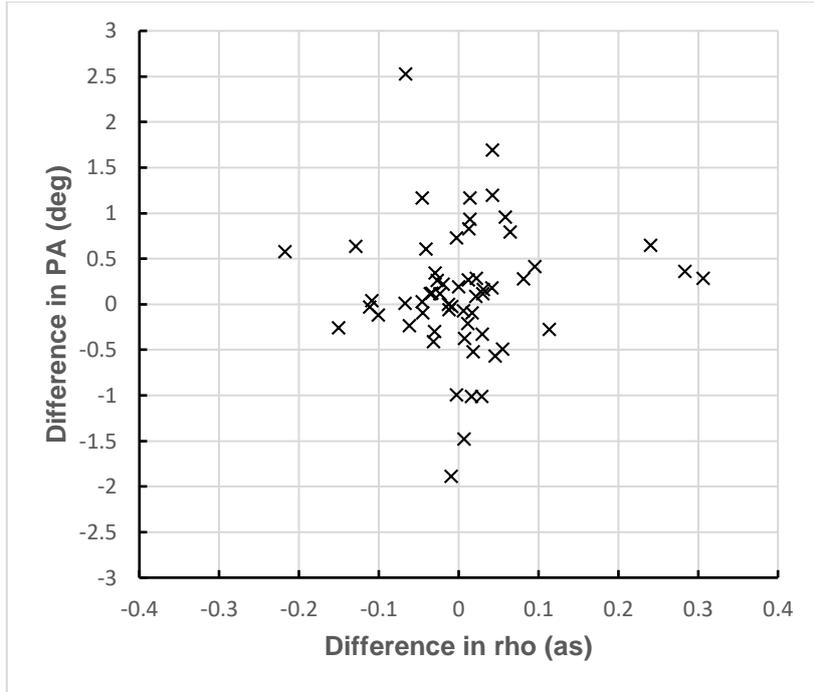

*Figure 4*: *The comparison of measures (GAIA DR2 ρ and PA – TW). The mean bias in ρ is 0.005 ± 0.021 arcseconds, and the mean bias in PA is 0.129 ± 0.094 degree.*

## 8.3 Comparison with Space-Based Positions.

The projected movement of the secondary relative to the primary shown in the rectilinear plots of Figure 5 as a dark line is the linear extrapolation of the HIPPARCOS and GAIA position at epoch 1991.25 and 2015.5 respectively. All positions shown in this figure are for equinox J2000.

Armed with this space-based differential proper motion, a comparison is made between our measures in Table 1 and the separations and PAs at the epoch of the observations (~2018.6).

This comparison gives a mean bias in separation of -0.032 ± 0.042 arcsec and a mean bias in PAs of -0.112 ± 0.081 degree. These biases again, are defined as TW minus space-based data.

## 8.4 Table of Comparisons.

Table 2 gives a summary of the bias as found by the three methods described above in Section 8. Our best estimate of bias in separations is ~10 mas, and ~40 millidegrees in PA.

Assuming that the uncertainties in the HIPPARCOS and GAIA data are negligible compared with our measures, we assert that our separations have a formal uncertainty of ~30 mas and the formal uncertainty in the PAs is ~80 millidegrees.

*Table 2: Comparison of Bias and Uncertainty.*

| BIAS COMPARISON | Sep (arcsec) | SEM Sep | PA (DEG) | SEM PA |
|---|---|---|---|---|
| Extrapolated Historic Data | -0.009 | 0.03 | 0.1 | 0.08 |
| GAIA DR2 | 0.005 | 0.02 | 0.13 | 0.09 |
| HIPPARCOS/GAIA Motion | -0.03 | 0.04 | -0.11 | 0.08 |
| Overall Estimate | -0.012 | 0.03 | 0.04 | 0.08 |

## 9. Conclusion.

We report lucky imaging measures of $\rho$ and PA of 62 southern pairs calibrated against both Alpha Centauri AB and drift scans and present them as final weighted measures. The formal uncertainties for the weighted measures are 0.080 arcsec in $\rho$ and 0.056 deg in PA.

We also report a comparison of our measures against extrapolated historic measures, GAIA DR2 and HIPPARCOSS/GAIA space-based measures. Our best estimate of bias within these three approaches are ~10 mas in $\rho$ and ~0.04 degrees in PA, and the formal uncertainty is ~30 mas and ~80 millidegrees in $\rho$ and PA respectively.

We present also Rectilinear Elements for the 61, Grade 5 Orbital Elements for 5 pairs and suggest 5 pairs as optical doubles.

**Acknowledgements**

- The Washington Double Star Catolog maintained by the USNO. (WDS), https://ad.usno.navy.mil/wds
- SIMBAD Astronomical Database, operated at CDS, Strasbourg, France, http://simbad.u-strasbg.fr/simbad/
- The Aladin sky atlas developed at CDS, Strasbourg Observatory, France, https://aladin.u-strasbg.fr/
- The Gaia Catalogue (Gaia DR2, Gaia Collaboration, 2018), from VizieR (GAIA DR2), http://vizier.u-strasbg.fr/viz-bin/VizieR-3?-source=I/345/gaia2
- The University of New England for the use of the Kirby Observatory.
- Thanks to Florent Losse for use of the *Reduc* software.
- Thanks to Jenny Stephens of Kildare Catholic College, Wagga Wagga.

**References**

Andrews, J. J., Chanamé, J & Agüeros, M. A., 2017, "Wide Binaries in Tycho-Gaia: search method and the distribution of orbital separations", Royal Astronomical Society, **472**, 675-699

Brosche, P., 1960, "Eine Bemerkung zur Massenbestimmung langperiodischer Doppelsternsysteme und Anwendung auf ads 10417", *Astronomische Nachrichten*, **285**(5), 261-264.

Brown, A.G.A., *et al*., 2018, "Gaia Data Release 2 - Summary of the contents and survey properties", *Astronomy and Astrophysics*, **616**, A1.

Fried, D. L., 1977, "Probability of getting a lucky short-exposure image through turbulence", *Journal of the Optical Society of America*, **68**(12), 1651-1658.

Irwin, A. W., Yang, S. L. S., & Walker, G. A. H., 1996, "36 Ophiuchi AB: Incompatibility of the orbit and precise Radial Velocities" *Astronomical Society of the Pacific*, **108**, 580-590.

James, M., Emery, M., White. G.L., Letchford, R.R. & Bosi, S.G., 2019a, "Measures of Ten Sco Doubles and the Determination of Two Orbits." *Journal of Double Star Observations,* **15**(3), 489-503.

James, M., 2019., BSc Honours Thesis, University of New England, Australia.

Letchford, R.R., White, G.L., & Ernest, A.D., 2018a, "The Southern Double Stars of Carl Rümker II: Their Relative Rectilinear Motion" *Journal of Double Star Observations*, **14**(2), 208–222.

Letchford, R.R., White, G.L., & Ernest, A.D., 2018b, "The Southern Double Stars of Carl Rümker III: Quantified Probability of Boundedness and Preliminary Grade 5 Orbits for Some Very Long Period Doubles." *Journal of Double Star Observations*, **14**(4), 761–770.

Mason, B.D., Wycoff, G.L., Hartkopf, W.I., Douglass, G.G., & Worley, C.E., 2001, "The 2001 US Naval Observatory double star CD-ROM. I. The Washington double star catalog." *Astron J*, **122**, 3466-3471.

MacEvoy, B., 2010, Private communication, http://www.handprint.com/ASTRO/index.html.

Nugent, R.L., & Iverson, E.W., 2011, "A New Video Method to Measure Double Stars", *Journal of Double Star Observations*, **7**(3),185-194.

Peirce, C. S., 1882, Harvard Ann, **13**, 17.

Pourbaix, D., & Boffin, H.M.J., 2016, *Astronomy and Astrophysics*, **586**, A90.

Prusti, T., *et al.*, (Gaia Collaboration)., 2016, The Gaia mission, *Astronomy & Astrophysics*, **595**, A1.

United States Naval Observatory., n.d., *The Washington Double Star Catalog*. Retrieved from http://ad.usno.navy.mil/wds/

White, G.L., Letchford, R.R. & Ernest, A.E., 2018, "Uncertainties in Separation and Position Angle of Historic Measures – Alpha Centarui AB Case Study", *Journal of Double Star Observations*, **14**(3), 432-442.

# Appendix

**Table A3**: *Measures at date of observation for this work (weighted α Cen AB orbital parameter and Video-drift measures), historics, GAIA DR2, and derived from Rectilinears (RE).*

| WDS | DISC | Rho | | | | PA | | | |
|---|---|---|---|---|---|---|---|---|---|
| | | This Work | History | GAIA | RE | This Work | History | GAIA | RE |
| 00582-1541 | S 390 | 6.452 | 6.406 | 6.464 | 6.419 | 216.290 | 216.139 | 216.077 | 215.523 |
| 01225-1905 | HJ 2043 | 5.115 | 5.127 | 5.179 | 5.187 | 72.134 | 73.224 | 72.926 | 73.641 |
| 01397-3728 | HJ 3452 | 19.610 | 19.713 | 19.655 | 20.059 | 277.769 | 277.133 | 277.203 | 276.283 |
| 01590-2255 | H 2 58AB | 8.711 | 8.607 | 8.682 | 8.295 | 302.246 | 302.021 | 302.589 | 304.213 |
| 02360-2124 | HJ 3511 | 14.702 | 14.577 | 14.720 | 14.786 | 98.232 | 98.264 | 97.714 | 97.863 |
| 02442-2530 | BSO 1AB | 12.529 | 12.569 | 12.493 | 12.102 | 192.518 | 193.143 | 192.634 | 189.987 |
| 02486-3724 | HJ 3532 | 5.320 | 5.246 | 5.341 | 5.429 | 144.265 | 143.734 | 144.353 | 145.758 |
| 03398-4022 | DUN 15 | 7.681 | 7.745 | 7.649 | 7.779 | 328.043 | 328.010 | 327.631 | 328.032 |
| 03486-3737 | DUN 16 | 8.409 | 8.411 | 8.400 | 7.634 | 217.178 | 217.291 | 215.291 | 210.494 |
| 13493-4031 | DUN 146 | 68.511 | 68.398 | 68.411 | 62.385 | 86.642 | 86.693 | 86.524 | 86.088 |
| 15036-2751 | HJ 4727AB | 7.346 | 7.343 | 7.362 | 7.534 | 221.429 | 221.506 | 220.420 | 218.209 |
| 15045-1754 | S 665 | 25.073 | 24.894 | 25.060 | 25.092 | 90.534 | 90.709 | 90.530 | 90.459 |
| 15103-4100 | COO 178 | 4.872 | 4.818 | 4.885 | 4.786 | 74.720 | 75.656 | 75.654 | 260.621 |
| 16086-3906 | DUN 199AC | 44.441 | 44.195 | 44.223 | 44.082 | 183.572 | 184.118 | 184.146 | 184.095 |
| 16482-3653 | DUN 209AB | 24.011 | 23.567 | 23.882 | 23.259 | 138.274 | 137.858 | 138.912 | 141.072 |
| 16540-4148 | JC 23AF | 56.702 | 56.603 | 56.656 | 56.659 | 19.114 | 20.204 | 20.280 | 20.305 |
| 17153-2636 | SHJ 243AB | 5.140 | 4.628 | 5.106 | 4.489 | 139.764 | 137.250 | 139.886 | 172.426 |
| 17180-2417 | H 3 25 | 9.988 | 10.230 | 10.001 | 10.431 | 353.349 | 354.375 | 354.174 | 354.157 |
| 17290-4358 | DUN 217 | 13.427 | 13.397 | 13.433 | 13.414 | 168.274 | 168.852 | 168.200 | 168.280 |
| 17317-4102 | ARY 114AC | 64.868 | 64.656 | 65.174 | 64.407 | 250.599 | 250.509 | 250.885 | 250.397 |
| 17398-0458 | STF2191AB | 26.350 | 26.392 | 26.238 | 26.514 | 266.705 | 267.220 | 266.674 | 266.895 |
| 17464-1318 | STF2204 | 14.509 | 14.511 | 14.463 | 14.269 | 24.027 | 24.465 | 24.056 | 24.643 |
| 18026-2415 | ARG 31AC | 35.600 | 35.189 | 35.532 | 35.414 | 26.296 | 26.550 | 26.308 | 26.731 |
| 18064-4145 | HJ 5011 | 28.143 | 28.217 | 28.185 | 29.022 | 344.074 | 344.325 | 345.267 | 347.112 |
| 18089-2528 | WNO 21 | 13.484 | 13.441 | 13.495 | 13.411 | 64.327 | 64.694 | 64.593 | 64.649 |
| 18106-1645 | S 700AB | 18.812 | 18.836 | 18.842 | 18.809 | 291.235 | 291.901 | 291.349 | 292.756 |
| 18106-1645 | S 700AC | 28.618 | 28.736 | 28.509 | 28.589 | 352.874 | 353.026 | 352.911 | 353.044 |
| 18108-4026 | HJ 5023 | 8.654 | 8.717 | 8.613 | 8.760 | 275.919 | 276.042 | 276.525 | 275.003 |
| 18187-1837 | SHJ 264AB,C | 17.230 | 17.388 | 17.244 | 17.261 | 51.048 | 51.342 | 52.216 | 50.442 |
| 18247-0636 | STF2313 | 5.762 | 6.011 | 5.857 | 6.047 | 195.693 | 196.396 | 196.105 | 196.959 |
| 18281-2645 | ARY 126AC | 54.054 | 54.067 | 54.095 | 54.284 | 135.204 | 134.849 | 135.383 | 135.476 |
| 18290-2635 | WNO 6 | 41.804 | 41.860 | 41.917 | 41.959 | 182.311 | 181.977 | 182.037 | 181.992 |
| 18334-3844 | DUN 222 | 20.470 | 21.322 | 21.344 | 21.376 | 358.654 | 359.032 | 358.418 | 357.954 |
| 18399-2531 | ARG 32AB | 7.032 | 7.382 | 7.049 | 7.135 | 219.503 | 219.013 | 219.409 | 217.689 |
| 18399-2531 | ARG 32AC | 53.217 | 53.363 | 53.299 | 53.296 | 284.941 | 285.436 | 285.222 | 285.108 |
| 18459-1030 | STF2373 | 4.117 | 4.151 | 4.125 | 4.164 | 337.020 | 337.369 | 336.642 | 336.416 |
| 19006-0807 | STF2425 | 29.534 | 29.562 | 29.592 | 30.640 | 177.100 | 177.819 | 178.055 | 179.455 |
| 19011-3704 | BSO 14AB | 12.825 | 12.936 | 12.822 | 12.736 | 280.033 | 280.386 | 280.763 | 280.977 |
| 19027-3606 | HJ 5080 | 5.484 | 5.541 | 5.472 | 5.526 | 245.883 | 246.013 | 245.822 | 245.306 |
| 19050-0402 | SHJ 286 | 39.705 | 39.620 | 39.555 | 38.331 | 209.937 | 210.092 | 209.679 | 208.888 |
| 19127-3351 | HJ 5094AB | 31.571 | 29.774 | 31.505 | 22.335 | 181.613 | 178.655 | 184.141 | 192.272 |
| 19177-1558 | S 715 | 8.380 | 8.442 | 8.387 | 8.423 | 16.967 | 16.555 | 15.491 | 14.437 |
| 19181-1557 | S 716 | 4.969 | 5.063 | 4.961 | 4.948 | 196.095 | 195.695 | 196.069 | 195.594 |
| 19431-0818 | STF45AB | 96.268 | 96.489 | 96.551 | 97.083 | 145.488 | 146.183 | 145.847 | 146.015 |
| 20178-4011 | DUN 230 | 9.675 | 9.753 | 9.645 | 9.825 | 117.438 | 117.325 | 117.138 | 116.131 |
| 20205-2912 | HJ 5188AC | 27.266 | 27.099 | 27.246 | 27.303 | 320.898 | 320.902 | 321.117 | 321.212 |
| 20284-1309 | STF2683 | 22.795 | 22.763 | 22.750 | 23.000 | 66.820 | 67.105 | 66.724 | 66.975 |
| 20299-1835 | SHJ 324 | 21.947 | 21.934 | 21.923 | 21.882 | 238.443 | 238.666 | 238.557 | 238.450 |
| 20322-2209 | HJ 2973AB | 39.386 | 39.107 | 39.417 | 39.385 | 128.953 | 128.795 | 129.115 | 129.105 |
| 20338-4033 | JC 18AB | 4.413 | 4.522 | 4.413 | 4.424 | 223.593 | 223.604 | 223.782 | 224.421 |
| 20484-1812 | S 763AB | 15.605 | 15.771 | 15.543 | 15.583 | 293.786 | 293.560 | 293.551 | 294.044 |
| 20501-2722 | HJ 5226 | 18.658 | 18.626 | 18.680 | 18.626 | 67.357 | 67.855 | 67.641 | 67.609 |
| 21022-4300 | DUN 236 | 57.422 | 57.386 | 57.451 | 57.433 | 73.193 | 72.564 | 72.865 | 72.747 |
| 22246-4127 | JC 19AB | 14.790 | 14.327 | 15.035 | 25.084 | 58.811 | 64.538 | 65.164 | 74.041 |

| WDS | DISC | | | | | | | |
|---|---|---|---|---|---|---|---|---|
| 22258-2014 | S 808AB | 6.869 | 7.240 | 6.898 | 6.855 | 152.910 | 152.627 | 151.897 | 150.871 |
| 22305-0807 | STF2913 | 8.067 | 8.125 | 8.040 | 8.090 | 327.936 | 328.252 | 328.200 | 328.625 |
| 23141-0855 | STF2993AB | 25.068 | 25.137 | 25.110 | 25.308 | 175.134 | 175.668 | 176.826 | 175.974 |
| 23141-0855 | S 826AC | 79.127 | 77.289 | 78.566 | 105.377 | 131.875 | 130.493 | 127.883 | 119.657 |
| 23238-0828 | STF3008 | 7.053 | 4.384 | 7.064 | 3.425 | 146.790 | 140.137 | 153.042 | 218.276 |
| 23460-1841 | H 2 | 6.950 | 7.008 | 7.005 | 6.424 | 135.798 | 134.898 | 135.305 | 136.728 |
| 23544-2703 | LAL 192 | 6.455 | 6.424 | 6.452 | 6.647 | 272.184 | 271.023 | 271.190 | 269.913 |
| 23595-2631 | LAL 193 | 10.373 | 10.155 | 10.613 | 10.565 | 169.436 | 169.937 | 170.084 | 170.184 |

***Table A4**: Difference in the measures determined for this work (TW), hisorical data (HIST), and GAIA DR2 (GAIA).*

| WDS | DISC | Diff PA | Diff Rho | Diff PA | Diff Rho | Diff PA | Diff Rho |
|---|---|---|---|---|---|---|---|
| | | GAIA-TW | | HIST-TW | | HIST-GAIA | |
| 00582-1541 | S 390 | -0.213 | 0.012 | -0.151 | -0.046 | 0.062 | -0.058 |
| 01225-1905 | HJ 2043 | 0.792 | 0.064 | 1.090 | 0.012 | 0.298 | -0.052 |
| 01397-3728 | HJ 3452 | -0.567 | 0.045 | -0.636 | 0.103 | -0.069 | 0.058 |
| 01590-2255 | H 2 58AB | 0.343 | -0.030 | -0.225 | -0.105 | -0.568 | -0.075 |
| 02360-2124 | HJ 3511 | -0.519 | 0.018 | 0.032 | -0.124 | 0.550 | -0.142 |
| 02442-2530 | BSO 1AB | 0.116 | -0.036 | 0.625 | 0.040 | 0.509 | 0.076 |
| 02486-3724 | HJ 3532 | 0.088 | 0.021 | -0.531 | -0.074 | -0.619 | -0.095 |
| 03398-4022 | DUN 15 | -0.412 | -0.031 | -0.033 | 0.064 | 0.379 | 0.096 |
| 03486-3737 | DUN 16 | -1.887 | -0.010 | 0.113 | 0.001 | 2.000 | 0.011 |
| 13493-4031 | DUN 146 | -0.118 | -0.101 | 0.051 | -0.113 | 0.168 | -0.012 |
| 15036-2751 | HJ 4727AB | -1.009 | 0.016 | 0.077 | -0.003 | 1.087 | -0.019 |
| 15045-1754 | S 665 | -0.004 | -0.013 | 0.175 | -0.178 | 0.178 | -0.166 |
| 15103-4100 | COO 178 | 0.934 | 0.014 | 0.936 | -0.054 | 0.002 | -0.067 |
| 16086-3906 | DUN 199AC | 0.574 | -0.218 | 0.546 | -0.245 | -0.028 | -0.027 |
| 16482-3653 | DUN 209AB | 0.638 | -0.129 | -0.417 | -0.444 | -1.055 | -0.315 |
| 16540-4148 | JC 23AF | 1.166 | -0.046 | 1.090 | -0.099 | -0.076 | -0.053 |
| 17153-2636 | SHJ 243AB | 0.121 | -0.034 | -2.514 | -0.512 | -2.636 | -0.478 |
| 17180-2417 | H 3 25 | 0.825 | 0.013 | 1.026 | 0.242 | 0.201 | 0.229 |
| 17290-4358 | DUN 217 | -0.074 | 0.006 | 0.578 | -0.030 | 0.652 | -0.036 |
| 17317-4102 | ARY 114AC | 0.286 | 0.306 | -0.090 | -0.212 | -0.376 | -0.518 |
| 17398-0458 | STF2191AB | -0.031 | -0.111 | 0.515 | 0.042 | 0.546 | 0.154 |
| 17464-1318 | STF2204 | 0.029 | -0.046 | 0.438 | 0.002 | 0.409 | 0.048 |
| 18026-2415 | ARG 31AC | 0.012 | -0.067 | 0.254 | -0.411 | 0.242 | -0.343 |
| 18064-4145 | HJ 5011 | 1.193 | 0.042 | 0.251 | 0.074 | -0.942 | 0.032 |
| 18089-2528 | WNO 21 | 0.265 | 0.012 | 0.366 | -0.043 | 0.101 | -0.055 |
| 18106-1645 | S 700AB | 0.114 | 0.030 | 0.666 | 0.024 | 0.552 | -0.006 |
| 18106-1645 | S 700AC | 0.038 | -0.108 | 0.153 | 0.119 | 0.115 | 0.227 |
| 18108-4026 | HJ 5023 | 0.606 | -0.041 | 0.123 | 0.063 | -0.483 | 0.104 |
| 18187-1837 | SHJ 264AB,C | 1.168 | 0.014 | 0.295 | 0.158 | -0.873 | 0.144 |
| 18247-0636 | STF2313 | 0.412 | 0.095 | 0.704 | 0.249 | 0.291 | 0.154 |
| 18281-2645 | ARY 126AC | 0.179 | 0.042 | -0.355 | 0.013 | -0.534 | -0.029 |
| 18290-2635 | WNO 6 | -0.274 | 0.114 | -0.334 | 0.056 | -0.060 | -0.057 |
| 18334-3844 | DUN 222 | -0.235 | 0.874 | 0.379 | 0.853 | 0.614 | -0.022 |
| 18399-2531 | ARG 32AB | -0.094 | 0.017 | -0.489 | 0.349 | -0.395 | 0.333 |
| 18399-2531 | ARG 32AC | 0.281 | 0.082 | 0.496 | 0.146 | 0.215 | 0.064 |
| 18459-1030 | STF2373 | -0.378 | 0.007 | 0.349 | 0.034 | 0.726 | 0.026 |
| 19006-0807 | STF2425 | 0.955 | 0.058 | 0.719 | 0.028 | -0.236 | -0.031 |
| 19011-3704 | BSO 14AB | 0.730 | -0.003 | 0.353 | 0.111 | -0.377 | 0.114 |
| 19027-3606 | HJ 5080 | -0.061 | -0.012 | 0.131 | 0.056 | 0.191 | 0.068 |
| 19050-0402 | SHJ 286 | -0.257 | -0.151 | 0.155 | -0.086 | 0.413 | 0.065 |
| 19127-3351 | HJ 5094AB | 2.528 | -0.066 | -2.958 | -1.797 | -5.486 | -1.730 |
| 19177-1558 | S 715 | -1.476 | 0.007 | -0.412 | 0.062 | 1.064 | 0.055 |
| 19181-1557 | S 716 | -0.026 | -0.009 | -0.400 | 0.094 | -0.374 | 0.102 |
| 19431-0818 | STF45AB | 0.359 | 0.283 | 0.695 | 0.221 | 0.336 | -0.062 |
| 20178-4011 | DUN 230 | -0.300 | -0.030 | -0.113 | 0.079 | 0.187 | 0.109 |
| 20205-2912 | HJ 5188AC | 0.219 | -0.020 | 0.004 | -0.166 | -0.215 | -0.147 |
| 20284-1309 | STF2683 | -0.096 | -0.045 | 0.285 | -0.031 | 0.381 | 0.013 |
| 20299-1835 | SHJ 324 | 0.114 | -0.024 | 0.223 | -0.012 | 0.109 | 0.011 |

| WDS | DISC | | | | | |
|---|---|---|---|---|---|---|
| 20322-2209 | HJ 2973AB | 0.162 | 0.031 | -0.157 | -0.279 | -0.320 | -0.310 |
| 20338-4033 | JC 18AB | 0.189 | 0.000 | 0.012 | 0.109 | -0.178 | 0.109 |
| 20484-1812 | S 763AB | -0.235 | -0.062 | -0.226 | 0.165 | 0.009 | 0.227 |
| 20501-2722 | HJ 5226 | 0.285 | 0.022 | 0.498 | -0.032 | 0.214 | -0.054 |
| 21022-4300 | DUN 236 | -0.327 | 0.029 | -0.629 | -0.036 | -0.302 | -0.065 |
| 22246-4127 | JC 19AB | 6.353 | 0.246 | 5.727 | -0.462 | -0.626 | -0.708 |
| 22258-2014 | S 808AB | -1.013 | 0.029 | -0.283 | 0.371 | 0.730 | 0.342 |
| 22305-0807 | STF2913 | 0.264 | -0.027 | 0.316 | 0.058 | 0.052 | 0.085 |
| 23141-0855 | STF2993AB | 1.693 | 0.042 | 0.534 | 0.069 | -1.159 | 0.027 |
| 23141-0855 | S 826AC | -3.991 | -0.561 | -1.382 | -1.837 | 2.610 | -1.277 |
| 23238-0828 | STF3008 | 6.252 | 0.011 | -6.653 | -2.668 | -12.905 | -2.679 |
| 23460-1841 | H 2 | -0.493 | 0.055 | -0.901 | 0.057 | -0.407 | 0.002 |
| 23544-2703 | LAL 192 | -0.994 | -0.003 | -1.161 | -0.031 | -0.167 | -0.028 |
| 23595-2631 | LAL 193 | 0.647 | 0.240 | 0.501 | -0.218 | -0.146 | -0.458 |

***Table A5***: *Rectilinear Elements of the Secondary Component of the 62 Pairs*
*(In-depth details in Letchford, White and Ernest, 2018a)*

| WDS | DISC | x0 | xa | y0 | ya | t0 |
|---|---|---|---|---|---|---|
| | | ± | ± | ± | ± | ± |
| 00582-1541 | S 390 | -5.2244 | 0.0001 | -3.7296 | -0.0011 | 1940.7543 |
| | | 0.0118 | 0.0002 | 0.0130 | 0.0002 | 850.5143 |
| 01225-1905 | HJ 2043 | 1.4610 | 0.0012 | 4.9771 | -0.0005 | 1953.5703 |
| | | 0.0058 | 0.0001 | 0.0134 | 0.0002 | 920.5884 |
| 01397-3728 | HJ 3452 | 2.1952 | 0.0055 | -19.9383 | 0.0065 | 1948.9966 |
| | | 0.0052 | 0.0001 | 0.0298 | 0.0004 | 361.3999 |
| 01590-2255 | H 2 58AB | 4.6639 | -0.0005 | -6.8594 | -0.0056 | 1930.2237 |
| | | 0.0116 | 0.0001 | 0.0138 | 0.0002 | 80.1596 |
| 02360-2124 | HJ 3511 | -2.0226 | -0.0002 | 14.6465 | -0.0011 | 1953.6573 |
| | | 0.0067 | 0.0001 | 0.0342 | 0.0006 | 15506.0950 |
| 02442-2530 | BSO 1AB | -11.9185 | -0.0037 | -2.0988 | -0.0083 | 1942.0339 |
| | | 0.0144 | 0.0002 | 0.0492 | 0.0007 | 241.2325 |
| 02486-3724 | HJ 3532 | -4.4882 | 0.0024 | 3.0550 | 0.0013 | 1945.9391 |
| | | 0.0050 | 0.0001 | 0.0056 | 0.0001 | 186.6285 |
| 03398-4022 | DUN 15 | 6.5994 | -0.0010 | -4.1186 | 0.0008 | 1941.1690 |
| | | 0.0046 | 0.0001 | 0.0056 | 0.0001 | 958.2834 |
| 03486-3737 | DUN 16 | -6.5784 | -0.0017 | -3.8740 | -0.0137 | 1934.2705 |
| | | 0.0045 | 0.0001 | 0.0051 | 0.0001 | 18.6075 |
| 13493-4031 | DUN 146 | 4.2560 | -0.0040 | 62.2394 | 0.0893 | 1950.0286 |
| | | 0.0162 | 0.0002 | 0.0110 | 0.0002 | 6.5108 |
| 15036-2751 | HJ 4727AB | -5.9199 | 0.0054 | -4.6600 | -0.0027 | 1947.6650 |
| | | 0.0096 | 0.0001 | 0.0099 | 0.0001 | 140.8817 |
| 15045-1754 | S 665 | -0.2009 | -0.0003 | 25.0914 | -0.0005 | 1948.2751 |
| | | 0.0010 | 0.0000 | 0.0796 | 0.0012 | 186907.5167 |
| 16086-3906 | DUN 199AC | -43.9699 | -0.0020 | -3.1482 | -0.0011 | 1952.7770 |
| | | 0.0004 | 0.0000 | 0.0017 | 0.0000 | 201.3896 |
| 16482-3653 | DUN 209AB | -18.0940 | 0.0038 | 14.6144 | 0.0191 | 1949.5819 |
| | | 0.0304 | 0.0005 | 0.0302 | 0.0005 | 90.9595 |
| 16540-4148 | JC 23AF | 53.1379 | 0.0004 | 19.6613 | -0.0012 | 1952.6081 |
| | | 0.0123 | 0.0002 | 0.0098 | 0.0002 | 7042.2868 |
| 17153-2636 | SHJ 243AB | -4.4500 | 0.0066 | 0.5917 | 0.0334 | 1937.6418 |
| | | 0.0050 | 0.0001 | 0.0063 | 0.0001 | 11.3784 |
| 17180-2417 | H 3 25 | 10.3771 | -0.0045 | -1.0620 | 0.0006 | 1924.7128 |
| | | 0.0039 | 0.0000 | 0.0269 | 0.0003 | 119.0806 |
| 17290-4358 | DUN 217 | -13.1344 | -0.0003 | 2.7248 | 0.0004 | 1949.9603 |
| | | 0.0059 | 0.0001 | 0.0199 | 0.0003 | 17131.8812 |
| 17317-4102 | ARY 114AC | -21.6092 | 0.0094 | -60.6742 | -0.0379 | 1991.7989 |
| | | 0.0091 | 0.0004 | 0.0075 | 0.0003 | 22.0390 |
| 17398-0458 | STF2191AB | -1.4360 | 0.0004 | -26.4753 | 0.0027 | 1923.9121 |
| | | 0.0023 | 0.0001 | 0.0293 | 0.0003 | 3006.9364 |
| 17464-1318 | STF2204 | 12.9694 | 0.0024 | 5.9496 | 0.0000 | 1922.9025 |
| | | 0.0356 | 0.0004 | 0.0303 | 0.0003 | 1265.1774 |
| 18026-2415 | ARG 31AC | 31.6294 | 0.0035 | 15.9291 | -0.0021 | 1959.9759 |
| | | 0.2942 | 0.0053 | 0.2845 | 0.0051 | 12043.6334 |

| WDS | Name | | | | | |
|---|---|---|---|---|---|---|
| 18064-4145 | HJ 5011 | 28.2907 | -0.0186 | -6.4731 | -0.0181 | 1957.0944 |
| | | 0.0254 | 0.0004 | 0.0167 | 0.0003 | 89.8055 |
| 18089-2528 | WNO 21 | 5.7420 | 0.0012 | 12.1192 | 0.0015 | 1971.4747 |
| | | 0.0199 | 0.0005 | 0.0304 | 0.0007 | 3392.0721 |
| 18106-1645 | S 700AB | 7.2755 | -0.0076 | -17.3448 | -0.0037 | 1960.9327 |
| | | 0.0564 | 0.0010 | 0.0494 | 0.0009 | 605.7845 |
| 18106-1645 | S 700AC | 28.3787 | -0.0014 | -3.4624 | -0.0008 | 1952.3457 |
| | | 0.0321 | 0.0005 | 0.0222 | 0.0004 | 11568.8688 |
| 18108-4026 | HJ 5023 | 0.7641 | 0.0022 | -8.7271 | 0.0020 | 1941.7357 |
| | | 0.0376 | 0.0005 | 0.0267 | 0.0004 | 1301.2426 |
| 18187-1837 | SHJ 264AB,C | 10.9929 | -0.0037 | 13.3077 | 0.0028 | 1948.4464 |
| | | 0.0211 | 0.0003 | 0.0208 | 0.0003 | 435.1843 |
| 18247-0636 | STF2313 | -5.7836 | 0.0018 | -1.7637 | 0.0019 | 1934.6681 |
| | | 0.0081 | 0.0001 | 0.0197 | 0.0002 | 622.6877 |
| 18281-2645 | ARY 126AC | -38.7020 | 0.0202 | 38.0648 | 0.0079 | 1996.4457 |
| | | 0.0109 | 0.0006 | 0.0109 | 0.0006 | 195.0099 |
| 18290-2635 | WNO 6 | -41.9332 | 0.0011 | -1.4584 | -0.0006 | 1970.5975 |
| | | 0.0005 | 0.0000 | 0.0024 | 0.0001 | 1426.1575 |
| 18334-3844 | DUN 222 | 21.3622 | -0.0003 | -0.7631 | 0.0029 | 1943.0966 |
| | | 0.0025 | 0.0000 | 0.0599 | 0.0008 | 1875.3057 |
| 18399-2531 | ARG 32AB | -5.6460 | 0.0030 | -4.3619 | -0.0022 | 1941.4198 |
| | | 0.0115 | 0.0002 | 0.0119 | 0.0002 | 259.4075 |
| 18399-2531 | ARG 32AC | 13.8905 | 0.0028 | -51.4535 | 0.0007 | 1954.3728 |
| | | 0.0022 | 0.0000 | 0.0057 | 0.0001 | 584.6560 |
| 18459-1030 | STF2373 | 3.8166 | -0.0003 | -1.6662 | 0.0008 | 1932.5191 |
| | | 0.0114 | 0.0001 | 0.0191 | 0.0002 | 3084.2071 |
| 19006-0807 | STF2425 | -30.6390 | 0.0119 | 0.2915 | 0.0104 | 1928.8816 |
| | | 0.0609 | 0.0007 | 0.0433 | 0.0005 | 317.8516 |
| 19011-3704 | BSO 14AB | 2.4252 | -0.0016 | -12.5029 | -0.0014 | 1945.2068 |
| | | 0.0099 | 0.0001 | 0.0376 | 0.0005 | 1231.8330 |
| 19027-3606 | HJ 5080 | -2.3085 | 0.0009 | -5.0204 | 0.0004 | 1941.6415 |
| | | 0.0108 | 0.0001 | 0.0173 | 0.0002 | 2520.2942 |
| 19050-0402 | SHJ 286 | -33.5615 | -0.0115 | -18.5175 | -0.0209 | 1958.6131 |
| | | 0.0015 | 0.0000 | 0.0019 | 0.0000 | 3.4228 |
| 19127-3351 | HJ 5094AB | -21.8246 | -0.1405 | -4.7473 | 0.0530 | 1949.8259 |
| | | 0.0007 | 0.0000 | 0.0038 | 0.0001 | 0.7064 |
| 19177-1558 | S 715 | 8.1573 | -0.0019 | 2.1000 | 0.0040 | 1929.4843 |
| | | 0.0066 | 0.0001 | 0.0160 | 0.0002 | 205.9884 |
| 19181-1557 | S 716 | -4.7663 | 0.0000 | -1.3302 | -0.0006 | 1932.3242 |
| | | 0.0090 | 0.0001 | 0.0222 | 0.0003 | 1837.5219 |
| 19431-0818 | STF45AB | -80.4996 | 0.0074 | 54.2667 | 0.0007 | 1929.1407 |
| | | 0.0114 | 0.0001 | 0.0130 | 0.0002 | 510.0643 |
| 20178-4011 | DUN 230 | -4.3272 | -0.0017 | 8.8208 | -0.0036 | 1944.2837 |
| | | 0.0074 | 0.0001 | 0.0106 | 0.0001 | 302.2190 |
| 20205-2912 | HJ 5188AC | 21.2820 | -0.0016 | -17.1036 | -0.0004 | 1959.5577 |
| | | 0.0084 | 0.0002 | 0.0084 | 0.0001 | 2610.3672 |
| 20284-1309 | STF2683 | 8.9960 | -0.0008 | 21.1674 | -0.0028 | 1929.7784 |
| | | 0.0553 | 0.0006 | 0.0488 | 0.0006 | 4406.0174 |
| 20299-1835 | SHJ 324 | -11.4500 | 0.0002 | -18.6478 | -0.0008 | 1950.9119 |
| | | 0.0395 | 0.0006 | 0.0487 | 0.0008 | 34918.1703 |
| 20322-2209 | HJ 2973AB | -24.8421 | -0.0003 | 30.5625 | 0.0003 | 1951.7920 |
| | | 0.0025 | 0.0000 | 0.0026 | 0.0000 | 15621.2016 |
| 20338-4033 | JC 18AB | -3.1598 | -0.0004 | -3.0966 | 0.0007 | 1936.9804 |
| | | 0.0084 | 0.0001 | 0.0084 | 0.0001 | 1194.4817 |
| 20484-1812 | S 763AB | 6.3492 | -0.0014 | -14.2311 | -0.0001 | 1954.1102 |
| | | 0.0287 | 0.0005 | 0.0467 | 0.0008 | 7468.7603 |
| 20501-2722 | HJ 5226 | 7.0953 | 0.0000 | 17.2218 | 0.0010 | 1954.6303 |
| | | 0.0416 | 0.0007 | 0.0727 | 0.0012 | 36919.9427 |
| 21022-4300 | DUN 236 | 17.0342 | 0.0004 | 54.8487 | -0.0001 | 1950.8127 |
| | | 0.0013 | 0.0000 | 0.0026 | 0.0000 | 11913.1212 |
| 22246-4127 | JC 19AB | 6.8967 | 0.0100 | 24.1172 | -0.1643 | 1950.0529 |
| | | 0.0198 | 0.0003 | 0.0164 | 0.0003 | 7.7483 |
| 22258-2014 | S 808AB | -5.9882 | -0.0019 | 3.3369 | -0.0023 | 1946.3114 |
| | | 0.0081 | 0.0001 | 0.0111 | 0.0002 | 227.9759 |
| 22305-0807 | STF2913 | 6.9074 | -0.0010 | -4.2123 | -0.0005 | 1931.2844 |
| | | 0.0146 | 0.0002 | 0.0179 | 0.0002 | 2283.0812 |
| 23141-0855 | STF2993AB | -25.2456 | 0.0039 | 1.7769 | 0.0029 | 1940.1590 |

| | | 0.0035 | 0.0000 | 0.0306 | 0.0004 | 621.2004 |
|---|---|---|---|---|---|---|
| 23141-0855 | S 826AC | -52.1406 | -0.0048 | 91.5732 | -0.4773 | 1949.2814 |
| | | 0.0413 | 0.0006 | 0.0414 | 0.0006 | 6.7973 |
| 23238-0828 | STF3008 | -2.6887 | -0.0335 | -2.1216 | 0.0617 | 1921.9226 |
| | | 0.0080 | 0.0001 | 0.0133 | 0.0001 | 8.9509 |
| 23460-1841 | H 2 | -4.6773 | -0.0040 | 4.4033 | 0.0072 | 1945.1662 |
| | | 0.0079 | 0.0001 | 0.0079 | 0.0001 | 46.1036 |
| 23544-2703 | LAL 192 | -0.0101 | 0.0029 | -6.6467 | 0.0028 | 1948.3459 |
| | | 0.0005 | 0.0000 | 0.0122 | 0.0002 | 240.0132 |
| 23595-2631 | LAL 193 | -10.4105 | -0.0003 | 1.8012 | 0.0023 | 1958.9360 |
| | | 0.0046 | 0.0001 | 0.0177 | 0.0003 | 394.4704 |

***Figure A5****: Rectilinear Motion of 61 Pairs. 15103-4100 (COO 178) removed (Section 7). (Details of computational technique is given in Letchford, White and Ernest, 2018a)*

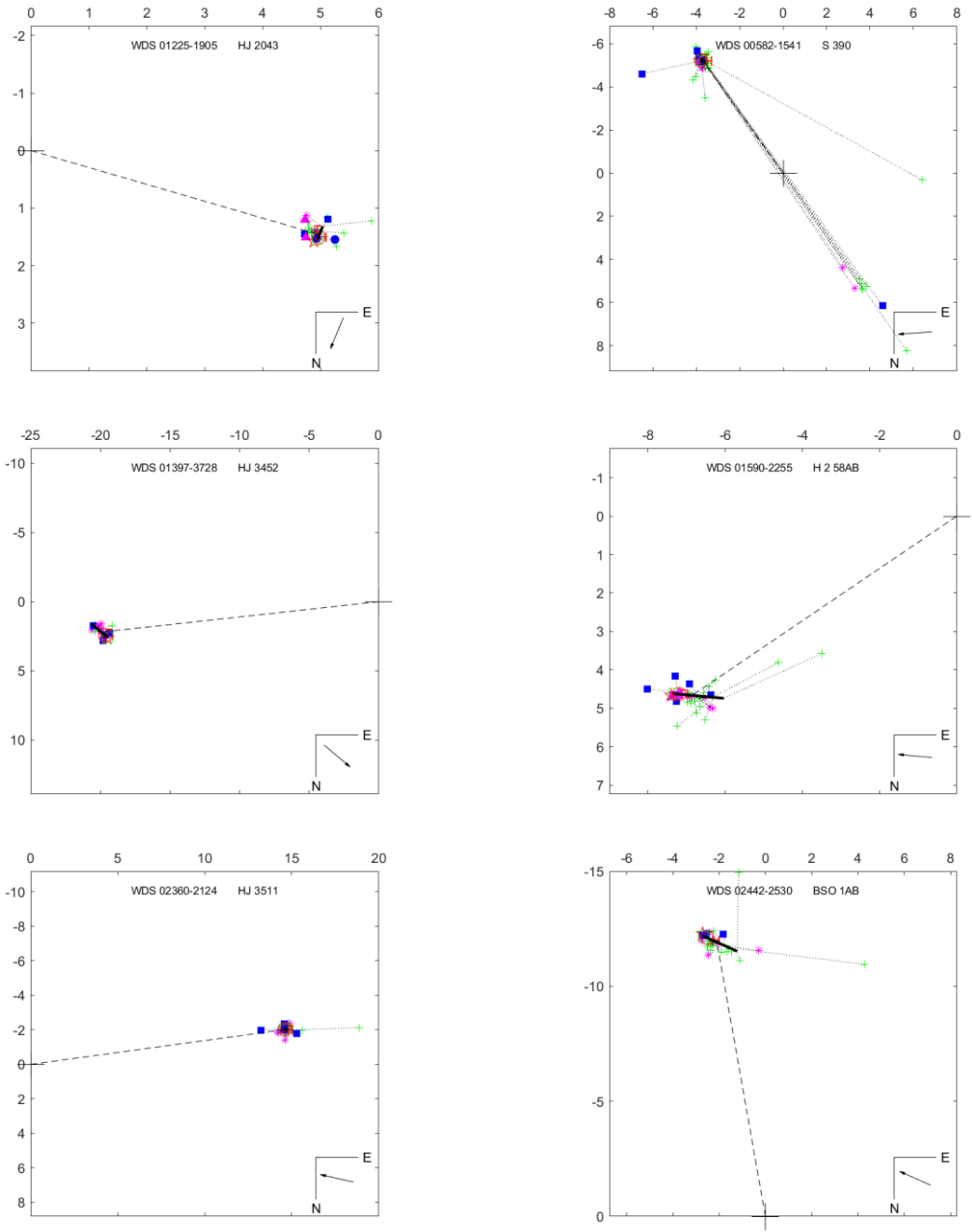

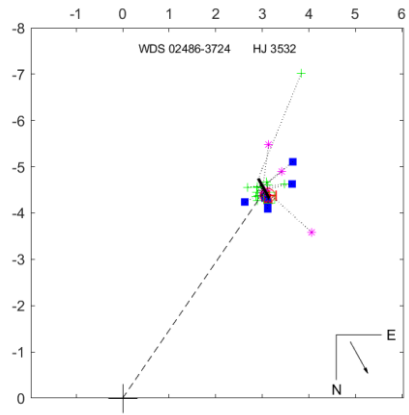
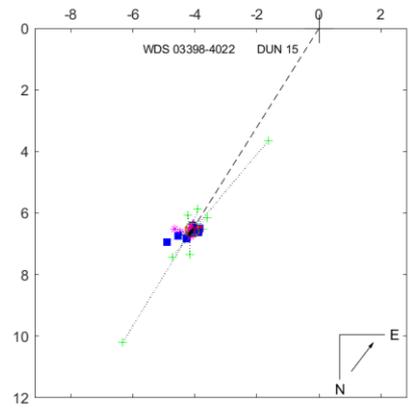
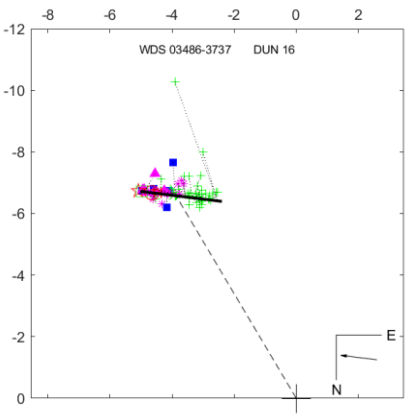
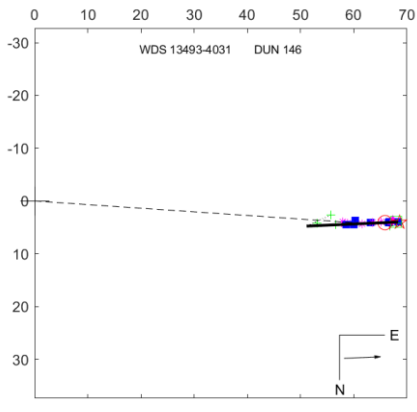
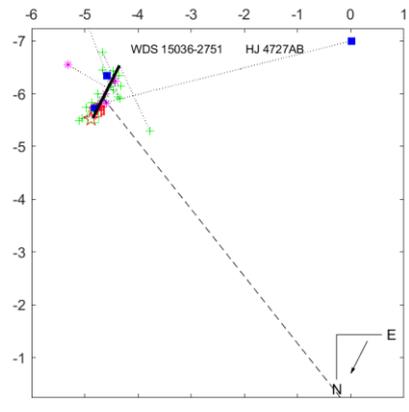
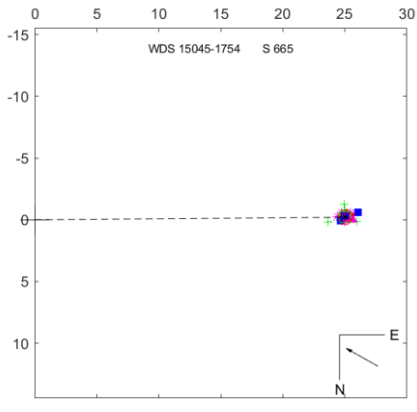
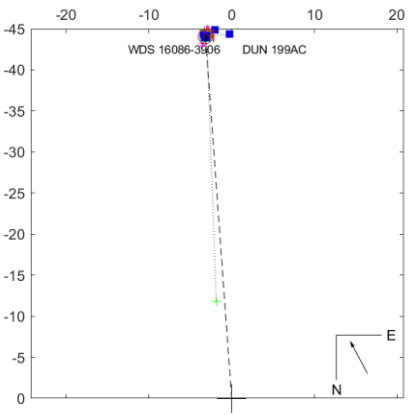
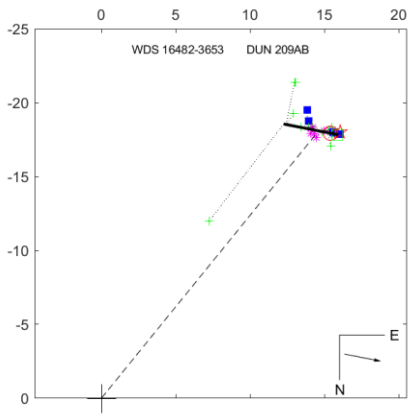

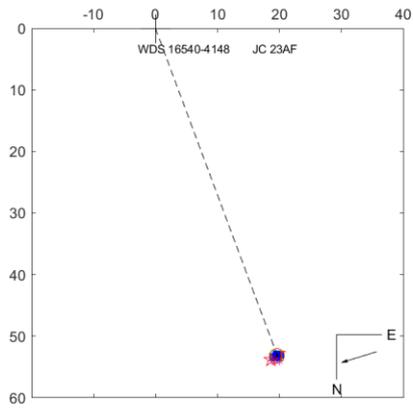
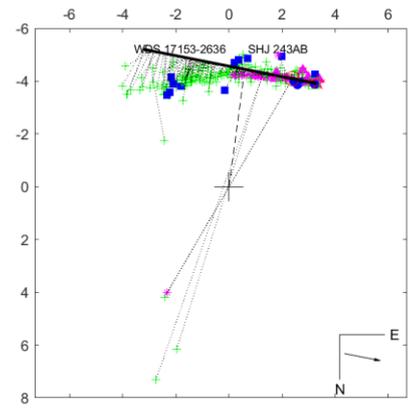
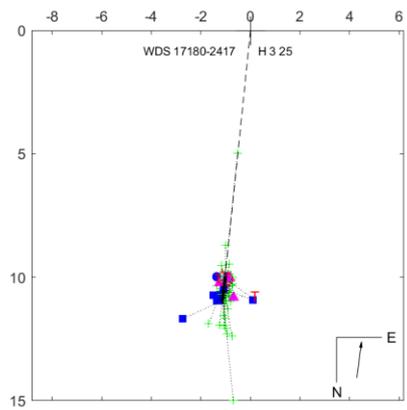
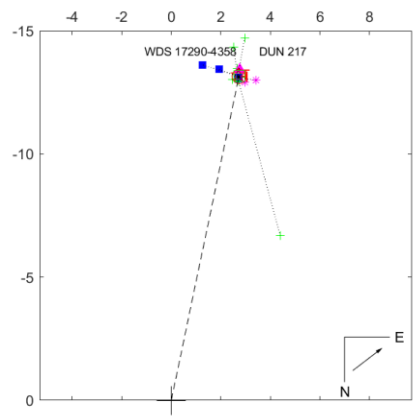
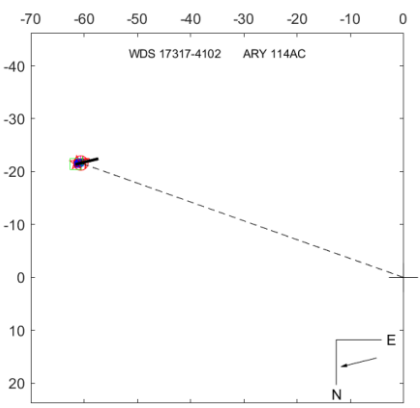
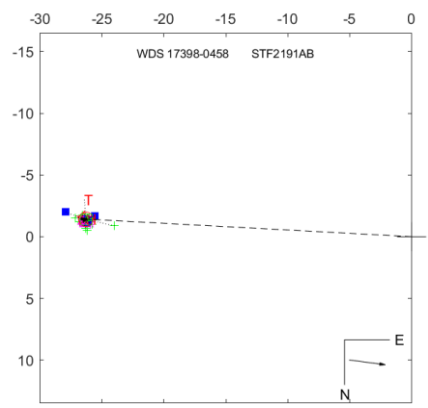
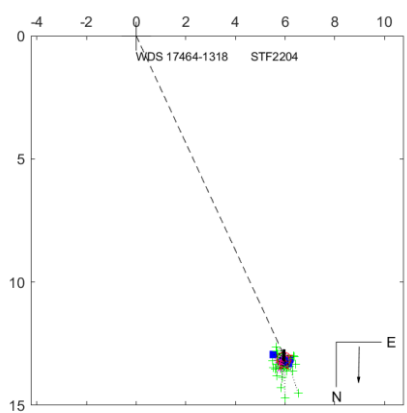
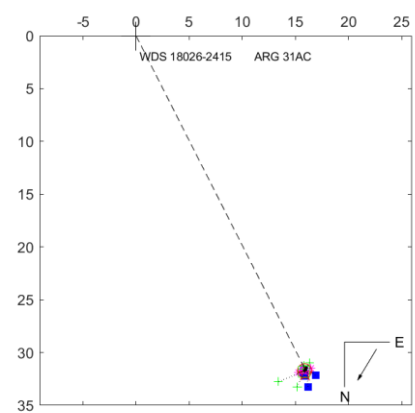

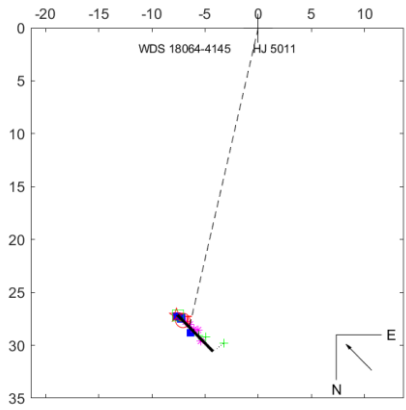
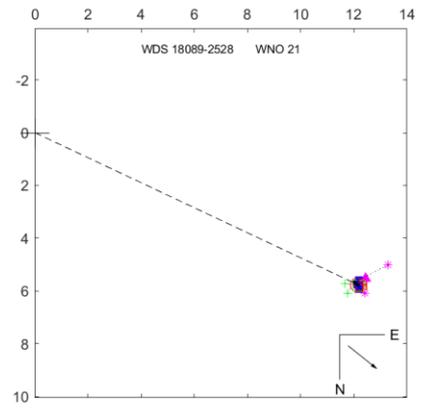
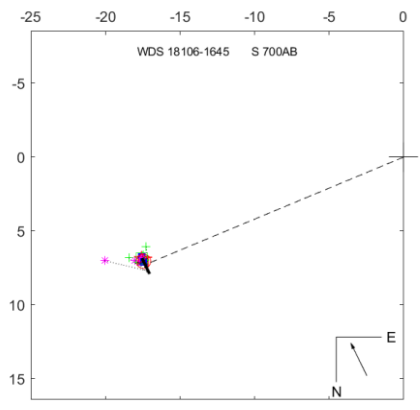
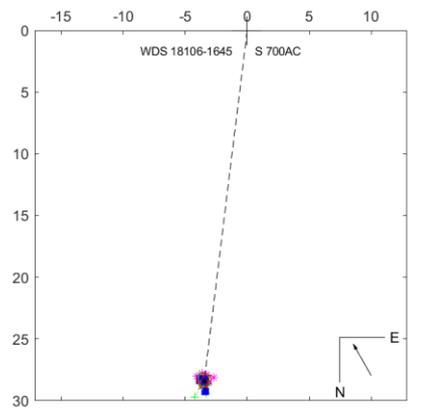
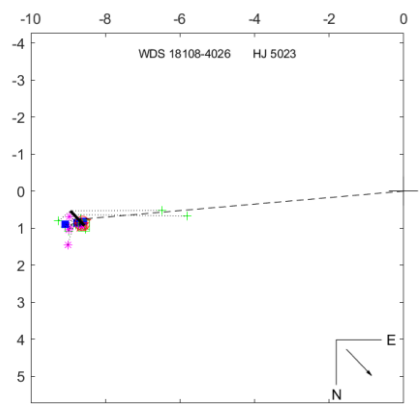
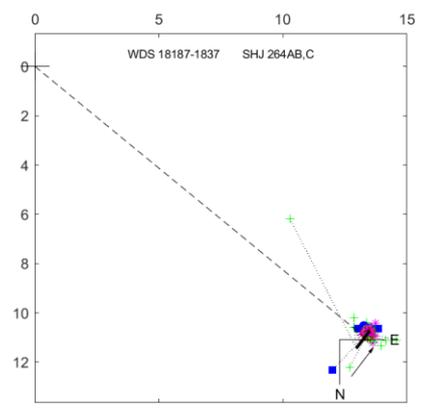
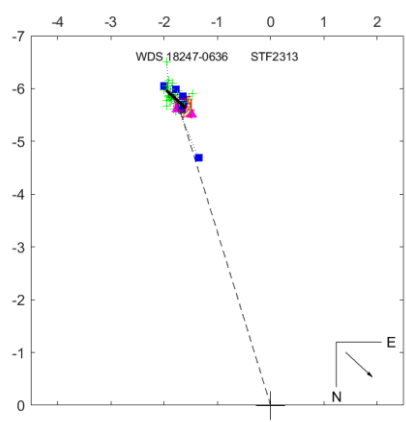
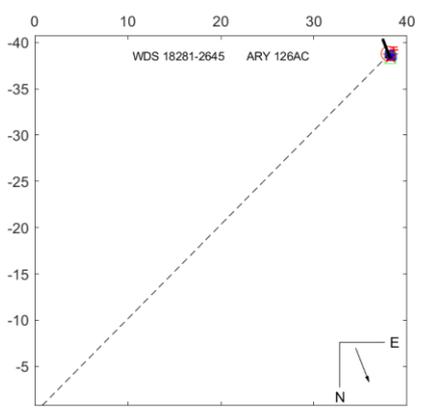

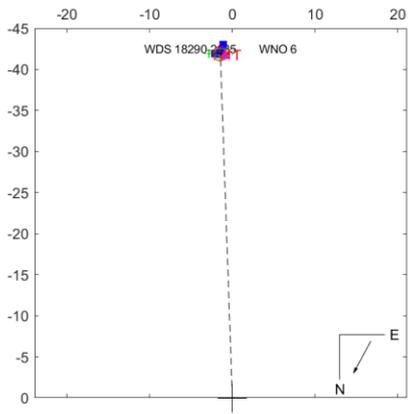
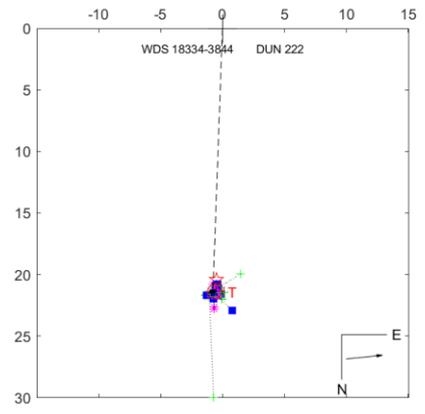
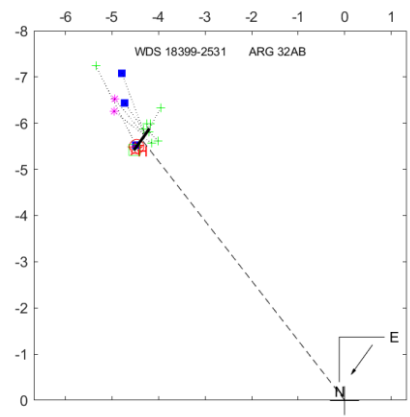
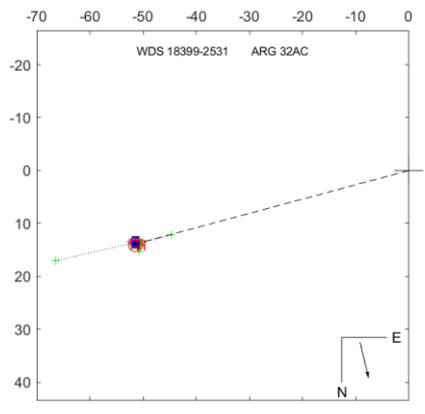
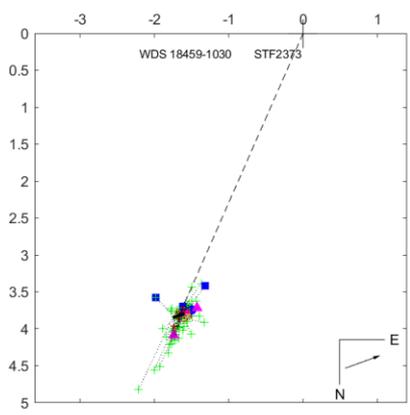
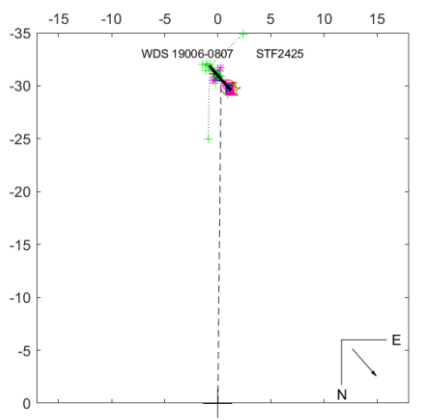
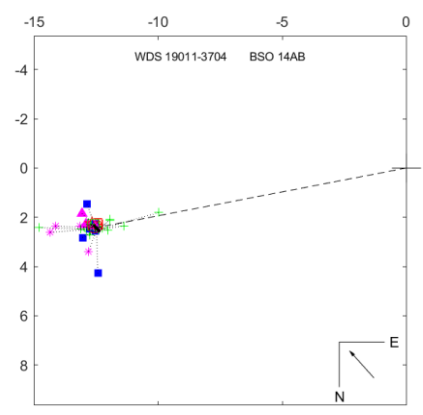
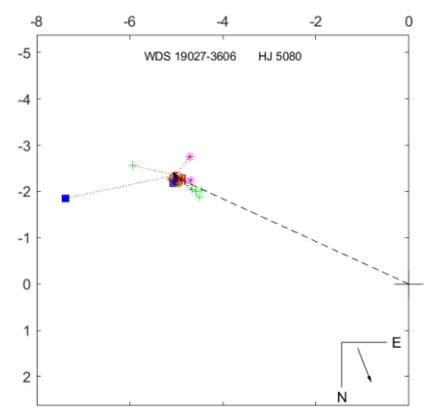

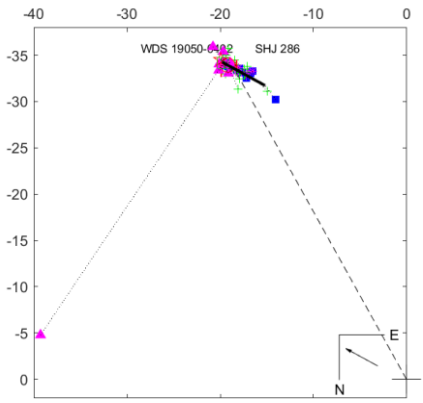
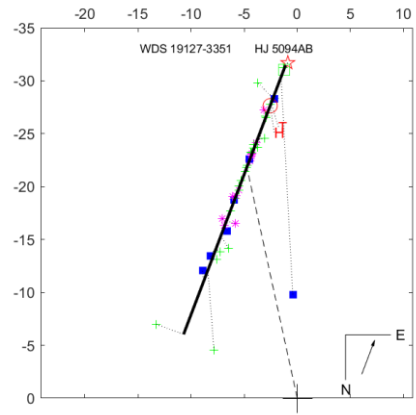
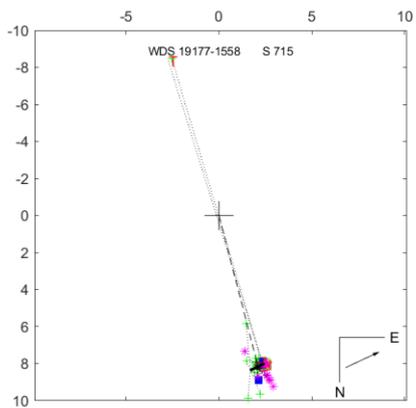
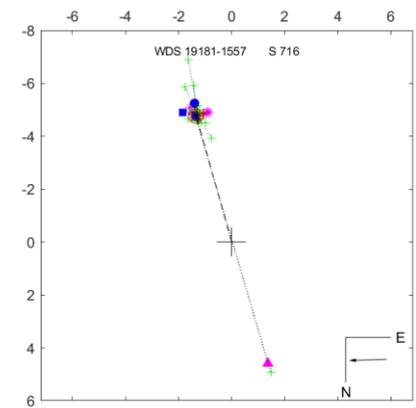
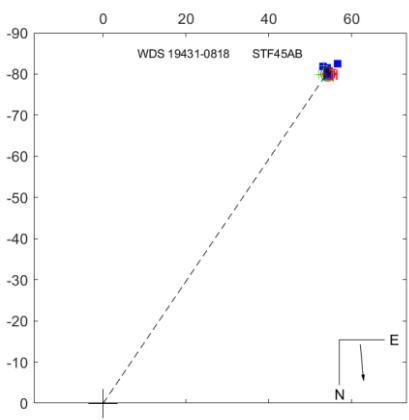
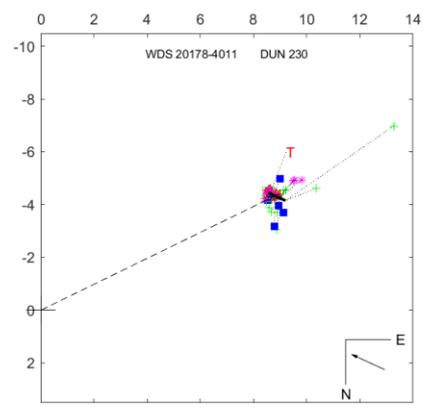
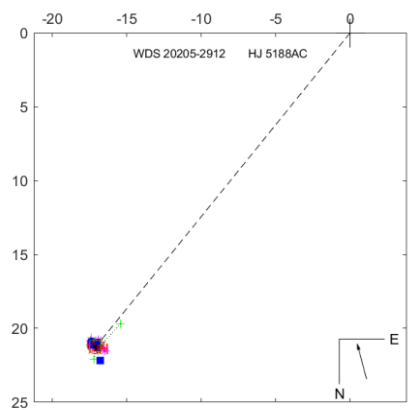
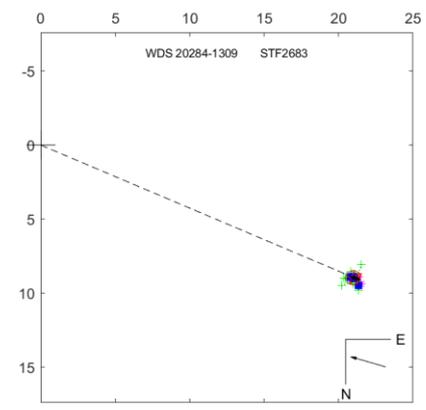

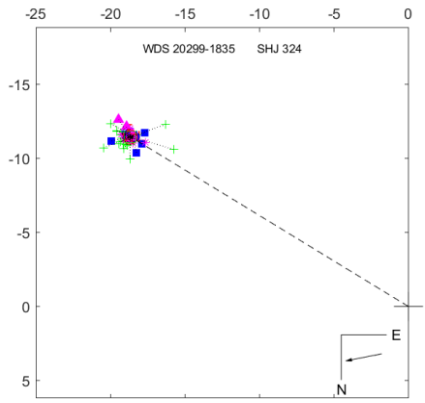
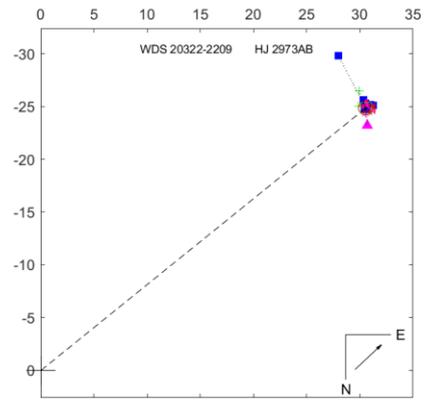
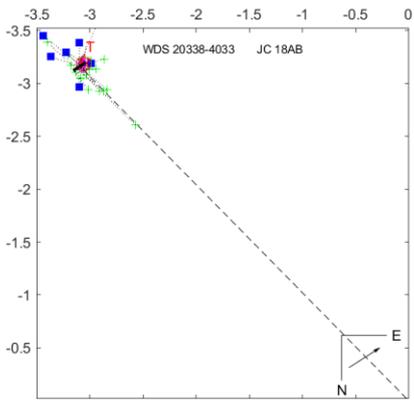
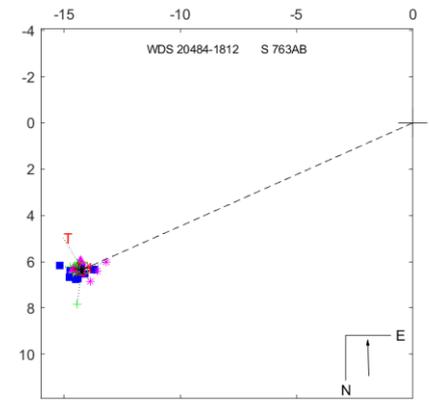
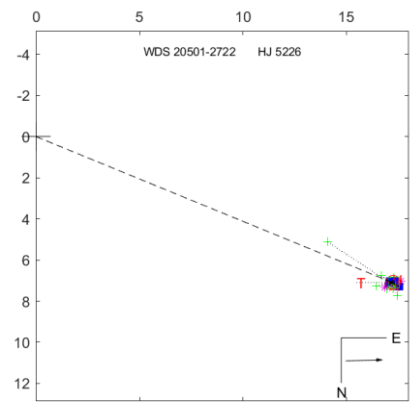
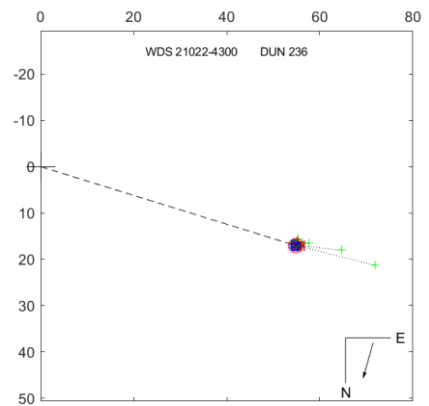
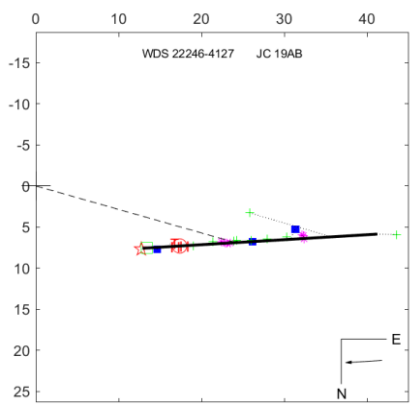
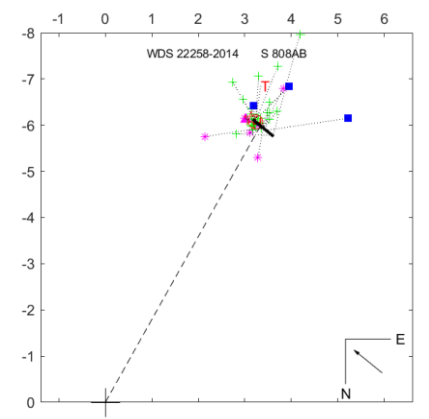

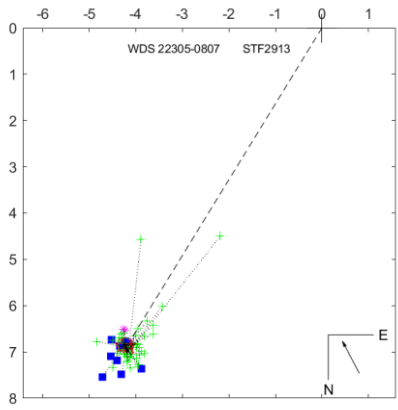
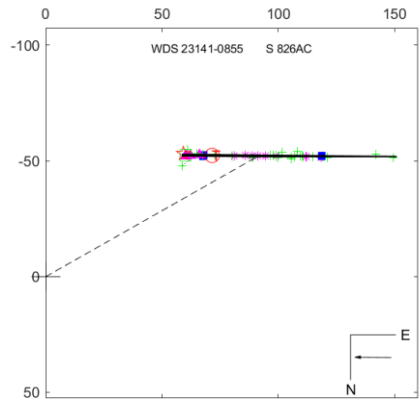
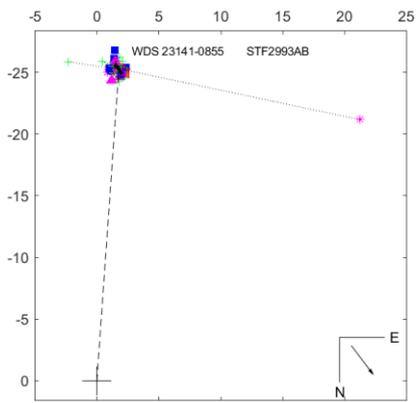
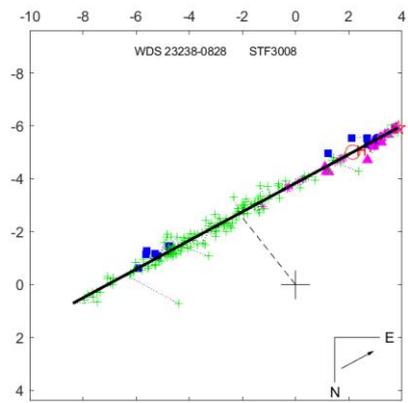
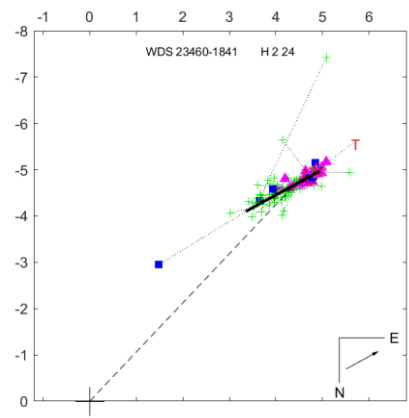
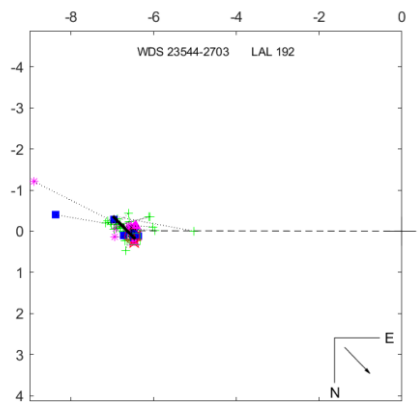
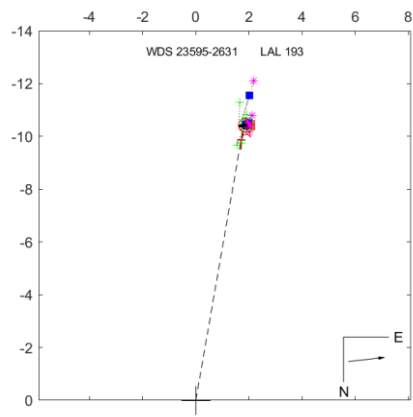

***Table A6***: *Grade 5 Orbital Elements for 5 Pairs.*
*(Details of computational technique is given in Letchford, White and Ernest, 2018b)*

| WDS | Disc | P yrs | a" | i° | Ω° | T yr | e | ω° |
|---|---|---|---|---|---|---|---|---|
| | | ± | ± | ± | ± | ± | ± | ± |
| 01590-2255 | H 2 58AB | 19551.176 | 12.760 | 169.176 | 179.714 | 700.866 | 0.820 | 108.335 |
| | | 3179.441 | 1.072 | 1.989 | 3.583 | 4.794 | 0.005 | 2.159 |
| 02360-2124 | HJ 3511 | 218347.189 | 30.564 | 64.094 | 66.945 | 16144.608 | 0.866 | 192.003 |
| | | 15408.740 | 1.497 | 1.202 | 2.473 | 1256.611 | 0.029 | 5.971 |
| 17153-2636 | SHJ 243AB | 507.062 | 12.490 | 100.386 | 87.635 | 2205.223 | 0.899 | 91.981 |
| | | 18.044 | 0.273 | 0.150 | 1.893 | 9.874 | 0.002 | 0.976 |
| 18247-0636 | STF2313 | 6685.981 | 6.473 | 104.775 | 38.582 | 5262.567 | 0.604 | 56.992 |
| | | 3551.475 | 1.515 | 1.617 | 14.506 | 915.520 | 0.046 | 22.403 |
| 23595-2631 | LAL 193 | 364086.947 | 71.700 | 119.721 | 163.699 | 2009.785 | 0.850 | 348.262 |
| | | 51528.580 | 5.789 | 1.196 | 5.522 | 4.595 | 0.012 | 6.091 |

**Figure A6**: Family of Orbits for the 5 Pairs. (Details of computational technique is given in Letchford, White and Ernest, 2018b)

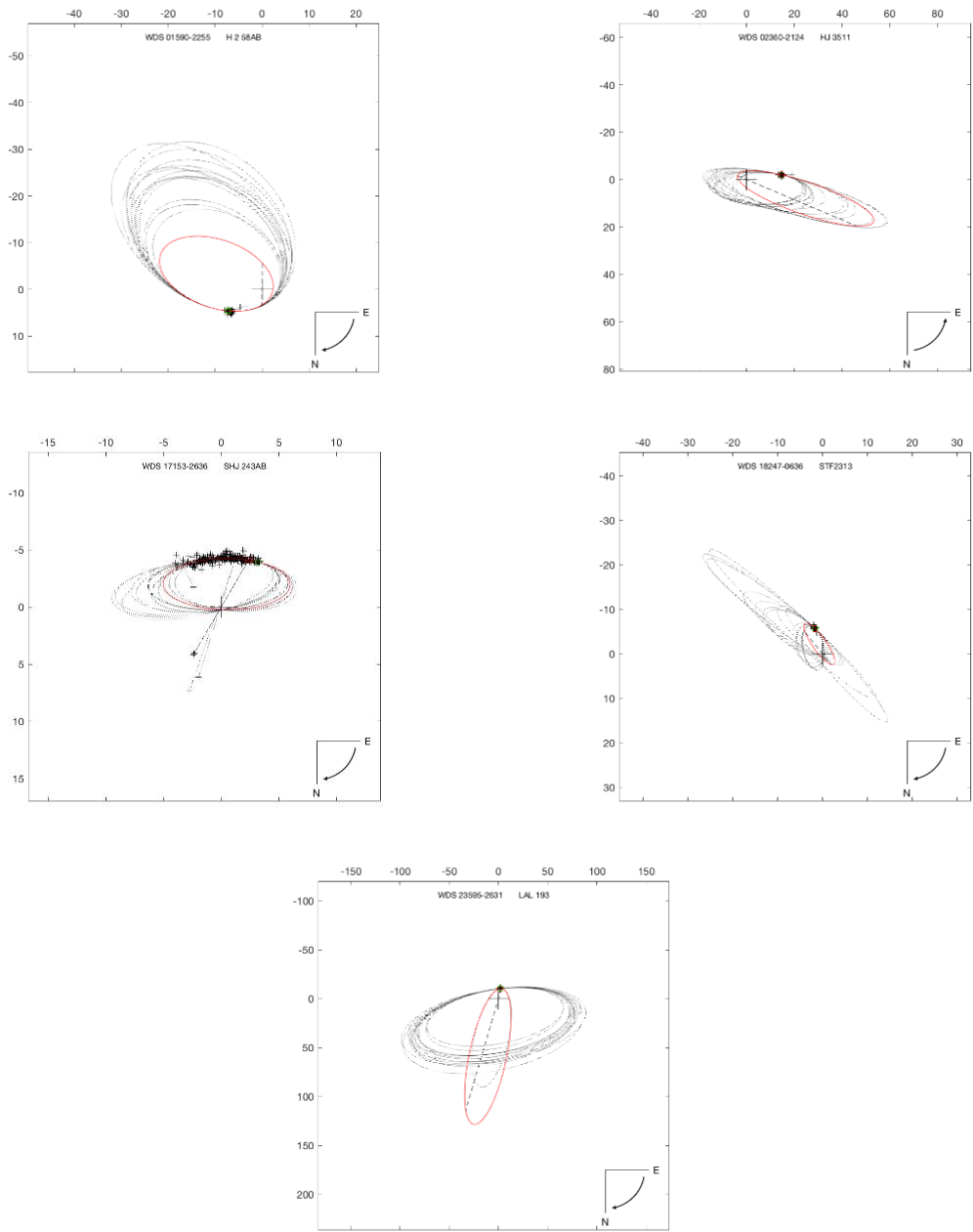